\documentclass[pdflatex, sn-nature,iicol, Numbered]{sn-jnl}% Default with double column layout
\usepackage{graphicx}%
\usepackage[final]{changes}
\usepackage{multirow}%
\usepackage{amsmath,amssymb,amsfonts}%
\usepackage{amsthm}%
\usepackage{mathrsfs}%
\usepackage[title]{appendix}%
\usepackage{xcolor}%
\usepackage{textcomp}%
\usepackage{manyfoot}%
\usepackage{booktabs}%
\usepackage{algorithm}%
\usepackage{algorithmicx}%
\usepackage{algpseudocode}%
\usepackage{placeins}
\usepackage{listings}%
\begin{document}

\title[Article Title]{From isolated polyelectrolytes to star-like assemblies:\\ The role of sequence heterogeneity on the statistical structure of the intrinsically disordered Neurofilament-low tail domain}

\author[1]{\fnm{Mathar} \sur{Kravikass}}

\author[1]{\fnm{Gil} \sur{Koren}}

\author[2]{\fnm{Omar A.} \sur{Saleh}}

\author*[1]{\fnm{Roy} \sur{Beck}}\email{roy@tauex.tau.ac.il}

\affil[1]{\orgdiv{School of Physics and Astronomy, the center for Nanoscience and Nanotechnology, and the center of Physics and Chemistry of Living Systems}, \orgname{Tel Aviv University}, \orgaddress{ \city{Tel Aviv}, \country{Israel}}}

\affil[2]{\orgdiv{Materials Department, Biomolecular Sciences and Engineering Program, and Physics Department}, \orgname{University of California}, \orgaddress{\city{Santa Barbara}, \country{USA}}}

\abstract{Intrinsically disordered proteins (IDPs) are a subset of proteins that lack stable secondary structure. Given their polymeric nature, previous mean-field approximations have been used to describe the statistical structure of IDPs. However, the amino-acid sequence heterogeneity and complex intermolecular interaction network have significantly impeded the ability to get proper approximations. One such case is the intrinsically disordered tail domain of Neurofilament low (NFLt), which comprises a 50 residue-long uncharged domain followed by a 96 residue-long negatively charged domain. Here, we measure two NFLt variants to identify the impact of the NFLt two main subdomains on its complex interactions and statistical structure. Using synchrotron small-angle x-ray scattering, we find that the uncharged domain of the NFLt induces attractive interactions that cause it to self-assemble into star-like polymer brushes. On the other hand, when the uncharged domain is truncated, the remaining charged N-terminal domains remain isolated in solution with typical polyelectrolyte characteristics. We further discuss how competing long- and short-ranged interactions within the polymer brushes dominate their ensemble structure, and, in turn, their implications on previously observed phenomena in NFL native and diseased states.

}

\keywords{Intrinsically disordered proteins, Small angle X-ray scattering, Neurofilament, Polymer physics, Polymer Brushes}

\maketitle
\section*{Introduction}
%Intro:
Intrinsically disordered proteins (IDPs) are a subset of proteins that, instead of forming a rigid singular structure, fluctuate between different conformations in their native form \cite{holehouse2023molecular,chowdhury2023interaction}. Nonetheless, IDPs serve significant biological functions and account for about 44\% of the human genome \cite{xue2012orderly}. The lack of fixed structure provides IDPs many advantages in regulatory systems in which they often play a crucial role in mediating protein interaction \cite{uversky2013intrinsic,ehm2022self}. These roles often come into play from intrinsically disordered regions (IDRs) of folded proteins interacting with other IDRs. For example, in the Neurofilament proteins, tails emanating from the self--assembled filament backbone domains bind together and form a network of filaments \cite{laser2015neurofilament,chernyatina2012atomic,malka2017phosphorylation, hirokawa1984organization, safinya2015assembly}. 

%IDP complex formation
The ensemble statistics of IDPs stem from their sequence composition and the surrounding solution \cite{chowdhury2023interaction}. For example, previous studies showed that IDPs comprising mostly negatively charged amino acids (polyelectrolytes) are locally stretched due to electrostatic repulsion between the monomers \cite{muller2010charge}. Moreover, different properties, such as hydrophobicity, were shown to be linked with local IDP domain collapse \cite{milles2012single}. The complex interactions that arise from sequence heterogeneity allow IDPs to form specific complexes without losing their disordered properties \cite{sekiyama2023toward}. For example, Khatun et al. recently showed how, under limited conditions, the human amylin protein self--assembles into fractal structures \cite{khatun2020fractal}.

%Analysis methods:
As IDPs are disordered chains, polymer theories are prime candidates to relate the measured structural statistics to known models, which can help link the sequence composition of the IDP to its conformations \cite{shea2021physics,van2014classification,baul2019sequence,das2013conformations}. Specifically, polymer scaling theories allow us to derive the statistical structure of IDPs given sequence--derived parameters, such as charge density and hydrophobicity \cite{muller2010charge,hofmann2012polymer,milles2012single,zheng2020hydropathy,maltseva2023fibril}. However, due to the heterogeneity of the IDP primary structure (\text{i.e.}, the amino acid sequence), some systems showed contradictions with the behavior theorized by standard heterogeneous polymer physics \cite{koren2023intramolecular, riback2017innovative, baul2019sequence,zeng2022competing, hofmann2012polymer}. 

The unique biological properties of IDPs have given rise to numerous attempts to use them as building blocks for self--assembled structures \cite{argudo2021folding}. For example, IDPs were proposed as brush--like surface modifiers, due to their enhanced structural plasticity to environmental conditions \cite{srinivasan2014stimuli,pregent2015probing}. Another example of an IDP brush system is the Neurofilament (NF) protein system \cite{laser2015neurofilament,beck2010gel,kornreich2016neurofilaments}, described as interacting bottle--brushes. NF subunit proteins form mature filaments with protruding disordered C--terminus IDR known as `tails.' NF tails were shown to mediate NF network formation and act as shock absorbents in high--stress conditions \cite{kornreich2016neurofilaments}. Moreover, NF aggregations are known to accumulate alongside other proteins in several neurodegenerative diseases, such as Alzheimer's,
Parkinson's, etc. \cite{didonna2019role}.  

The NF low disordered tail domain (NFLt) sequence can be divided into two unique regions: an uncharged region (residues 1--50) starting from its N terminal and a negatively charged region (residues 51--146). The NFLt can be described as a polyelectrolyte with a net charge per residue (NCPR) of -0.24. Furthermore, the statistical structures of segments within the NFLt are influenced by the amount, type, and disperse of the charged amino acid within a segment \cite{koren2023intramolecular}. Nonetheless, other structural constraints, particularly long--range contacts, impact the local statistical structures. Additionally, NFLt was shown to have glassy dynamics with the response to tension \cite{morgan2020glassy}. Such dynamics were associated with multiple weakly interacting domains and structural heterogeneity.

In this paper, we revisit NFLt as a model system for charged IDP and focus on the contribution of its neutral and hydrophobic N--terminal domain. We will show that increased salt concentration causes NFLt to form star--like brushes with increased aggregation number ($Z$). \added {Here, we are motivated by theoretical models, in particular the Pincus’ model for salted polyelectrolytes \cite{pincus1991colloid}, that capture key physical properties of IDPs, including the model system presented here \cite{srinivasan2014stimuli, kornreich2016neurofilaments, zhulina2009polyelectrolyte}}. We will further quantify the competition between hydrophobic attraction and electrostatic and steric repulsion in the formation of the structures of NFLt.

\newcommand{\dn}{$\rm{\Delta}$N42~}
\newcommand{\dnp}{$\rm{\Delta}$N42}

\section*{Results}

To study the N--terminal domain contribution to the structure of NFLt, we designed two variants and measured them at various buffer conditions. The first construct is the entire 146 residues of the NFLt chain, which we term as WT (NCPR = -0.24), and the second is isolating the 104 negatively charged residues from the C-terminal of NFLt (NCPR = -0.33), termed as \dnp. We expressed the variants in \textit{E-coli} and purified it up to 96\% (see methods).    

\begin{figure}[hbt!]\includegraphics[width=1\linewidth]{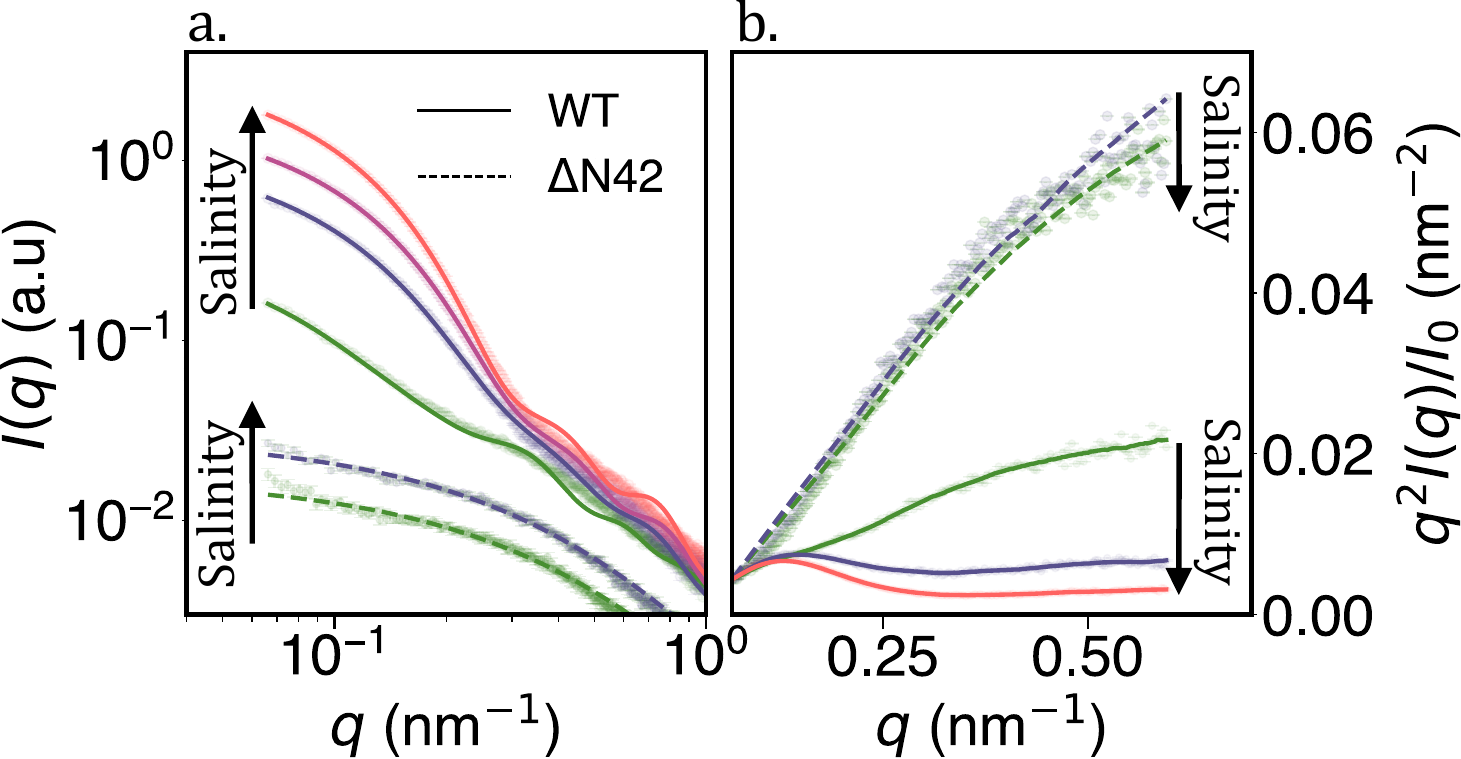}
\caption{SAXS measurements of WT and \dn at different salinity ($C_s$). \textbf{a.} For increasing $C_s$, the WT variant shows increased small angle scattering, a signature for aggregation. In contrast, \dn remains structurally intrinsically disordered as  $C_s$ vary. Data points are shifted for clarity. Lines are form-factor fittings, as described in the text. \textbf{b.} Normalized Kratky plot of the same SAXS measurements. The \dn variant remains disordered and unchanged with salinity, while the WT variant shows a hump at low $q$, typical for a collapse region. With increasing $C_s$, the hump at the lower $q$ range becomes a sharper peak accompanied by a scattering rise at the higher $q$ range. Such behavior indicates that the aggregation coexists with the WT variant's highly dynamic and disordered regions. Both variants shown are at the highest measured concentration (Table S1, S3). WT measurements are in 20 mM Tris pH 8.0 with 0, 150, 250, and 500 mM added NaCl (from bottom to top). Likewise, for \dnp, measurements are in 20 mM Tris pH 8.0 with 0 and 150 mM added NaCl (bottom to top).} 
\label{fig:raw_data}
\end{figure}

We assessed the variants in solution using small--angle X--ray scattering (SAXS), a technique extensively used to characterize the statistical structures of IDPs \cite{tria2015advanced}. From the raw SAXS data, measured at various salinities, we can already find high structural differences between the two variants (Fig. \ref{fig:raw_data}a). Dominantly at the low wave-vector ($q$) region, the WT variant scattering ($I$) rises with added NaCl salt. Such an increase at low $q$ implies high molecular mass particles due to aggregation of the WT variant. 

In contrast, \dn shows a separated Gaussian polymer profile (Figs. \ref{fig:raw_data}a, S1), nearly insensitive to total salinity ($C_s=20-520$~mM). Similarly, the data presented in Kratky format ($qI^2$ vs. $q$, Fig. \ref{fig:raw_data}a) shows the \dn has the signature of a disordered polymer. In contrast, the WT variant, in particular at high salinity, has a combination of a collapse domain (the peaks from below $q = 0.25 \rm{nm}^{-1}$) and a disordered polymeric structure (the scattering rise at higher $q$ Fig. \ref{fig:raw_data}a).    

Being completely disordered, \dn lacks a stable structure and can be described using a statistical ensemble of polymeric conformations \cite{zheng2018extended} were: 
\begin{equation}
\begin{split}
    I(q) = I_0 \exp\{ -\frac{1}{3}(qR_{\rm G})^2 + \\ 0.0479(\nu - 0.212)(qR_{\rm G})^4\}.
\end{split}
\label{eq:1}
\end{equation}
Here, $I_0$ is the scattering at $q=0$, $\nu$ is Flory scaling exponent, and $R_{\rm G}$ is the radius of gyration defined by:
\begin{equation}
    R_{\rm G} = \sqrt{\frac{\gamma(\gamma+1)}{2(\gamma+2\nu)(\gamma+2\nu+1)}}bN^\nu,
\label{eq:2}
\end{equation}
where $\gamma = 1.615$ and $b=0.55 \rm{nm}$ (see \cite{zheng2018extended}) and the analysis is viable up to $qR_{\rm G} \sim 2$ (Fig. S2, S3). In all \dn cases, the scattering profile fits Eq. \ref{eq:1} and with $\nu$ ranging between 0.63--0.69 depending on the buffer salinity (Table S1). In `infinite dilution' conditions (zero polymer concentration), we find $\nu$ to decrease monotonically from 0.73 to 0.62 with added salt (Table S2).    

Given the noticeable aggregation for the WT variant, alternative form factors were considered to match the scattering profiles (lines in Fig. \ref{fig:raw_data}). The absence of structural motifs at high $q$ values ($q > 0.3\,\, \rm{nm}^{-1}$) indicates a disordered nature for WT at shorter length scales. Conversely, in the lower $q$ region ($q < 0.3\,\, \rm{nm}^{-1}$), the scattering suggests stable structural motifs or a larger molecular weight particles. Such SAXS resembles that of self--assembled decorated spherical micelles \cite{pedersen2002scattering}. Variations of micelle models are shown to fit the data (Figs. \ref{fig:raw_data}, S4-S6). Sufficiently low aggregation number and core size distil the description of the spherical micelle into a `star--like' brush.  Alternative attempts to fit the scattering profiles to other form factors models, including vesicles and lamellar, were unsuccessful.

\begin{figure*}[ht!]
\centering
\includegraphics[width=1\linewidth]{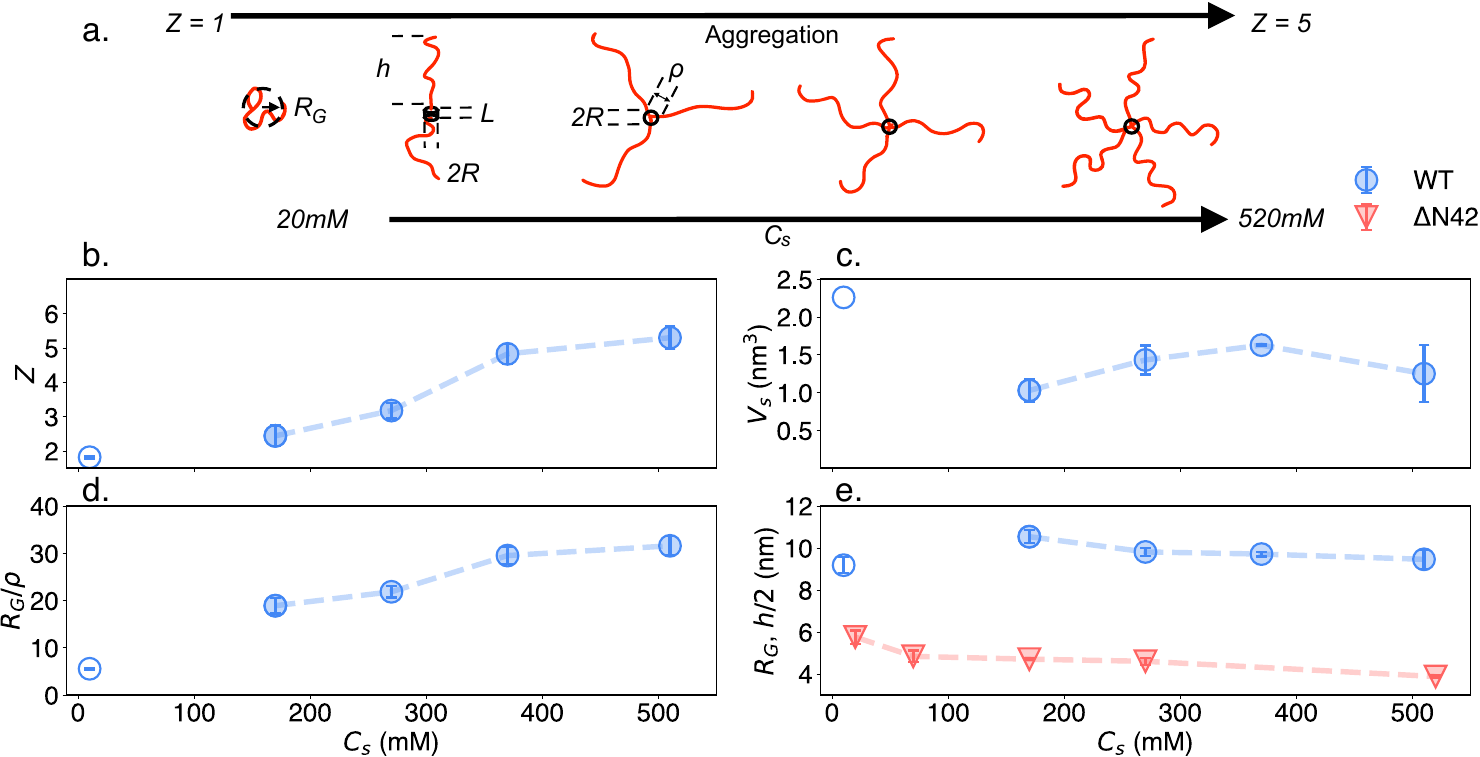}
\caption{\textbf{a.} Schematic of the system’s structure variation with salinity ($C_s$). While \dn remains disordered and segregated, the WT variant aggregates to a star--like polymer with a higher aggregation number at higher $C_s$. \textbf{b--e.} Structural parameters for WT (blue symbols) and \dn \replaced{(red symbols) }{(black symbols)} variants extracted from fitting the SAXS data. Full and hollow circles represent the spherical and cylindrical core fitted parameters, respectively. \textbf{d.} In all cases, the brush heights ($h$) are much larger than the corresponding grafting length ($\rho$), indicative of a brush regime. \textbf{e.} The structurally intrinsically disordered \dn variant compacts with higher $C_s$ values and remains more compacted from the projected brushes for the WT variant. All values are the extrapolated ‘zero concentration’ fitting parameters (see  Fig. S7)}
\label{fig:fitting}
\end{figure*}

For the star--like model, the aggregated variants form a small spherical core of volume $V_{\rm core}$ made out of $n \cdot Z$ monomers (comparison with different cores described in \cite{pedersen2000form} and in Fig. S4), where $n$ denotes the peptide length per polypeptide within the core, and $Z$ is the aggregation number, \text{i.e.} the number of polypeptides per `star.' The remainder of the WT variant then protrudes from the core as the star polymer brush (Figs. \ref{fig:fitting}a, S4-S6).
The star--like scattering form factor is described as a combination of four terms \cite{pedersen2002scattering}: the self--correlation term of the core $F_{\rm c}$, the self-correlation term of the tails $F_{\rm t}$, the cross-correlation term of the core and the tails $S_{\rm ct}$ and the cross-correlation term of the tails $S_{\rm tt}$:

\begin{equation}
    \begin{split}
        F_{\text{total}}(q) = Z^2 \beta_{\rm c}^2 F_{\rm c}(q) + Z \beta_{\rm t}^2 F_{\rm t}(q) +\\ 
        2Z^2 \beta_{\rm c} \beta_{\rm t} S_{\rm ct}(q) + Z(Z-1) \beta_{\rm t}^2 S_{\rm tt}(q).
    \end{split}
\label{eq:3}
\end{equation}

Here, $\beta_{\rm c}$ and $\beta_{\rm t}$ are the excess scattering length of the core and the tails, respectively. From fitting the scattering data, we extracted the height of the tails $h=2R_{\rm G}$, the aggregation number $Z$, and the relevant core's parameters (\textit{e.g.}, core radius $R$ for a spherical core, cylinder radius $R$ and length $L$ for a cylindrical core \cite{pedersen2000form}), schematically illustrated in Fig. \ref{fig:fitting}a. All fitting parameters are found in Table S3.

To avoid misinterpretation and to minimize intermolecular interaction effects, we present the fitting results at the `infinitely diluted regime' by extrapolating the relevant parameters measured at various protein concentrations to that at zero protein concentration (Fig. S7, Table S4). The parameters are mostly independent of the concentration unless explicitly mentioned.  

At low salinity (20mM), the aggregation number for the WT variant is of a dimer ($Z\approx 2$), and the core's shape is that of a cylinder (with a radius $R= 0.89$ nm and length $L=1.19$ nm). At higher salt conditions (170-520 mM), the form factor fits spherical core aggregates with increasingly higher $Z$'s (Fig. \ref{fig:fitting}a).  

Given the relatively small core volume ($V_{\rm core}\approx 1-2 \rm{nm}^3$, Fig. \ref{fig:fitting}c), it is crucial to evaluate the `grafting' distance between neighboring chains, $\rho$, on the core surface ($S=4\pi R^2 = Z\rho^2$) and the brush extension, $h$,  outside the core. As shown in Fig. \ref{fig:fitting}d, in all cases, $h/\rho \gg 1$ indicates a `brush regime' where neighboring chains repel each other while extending the tail's height \cite{chen201750th}. 

\begin{figure*}[hbt!]
\centering
\includegraphics[width=0.7\linewidth]{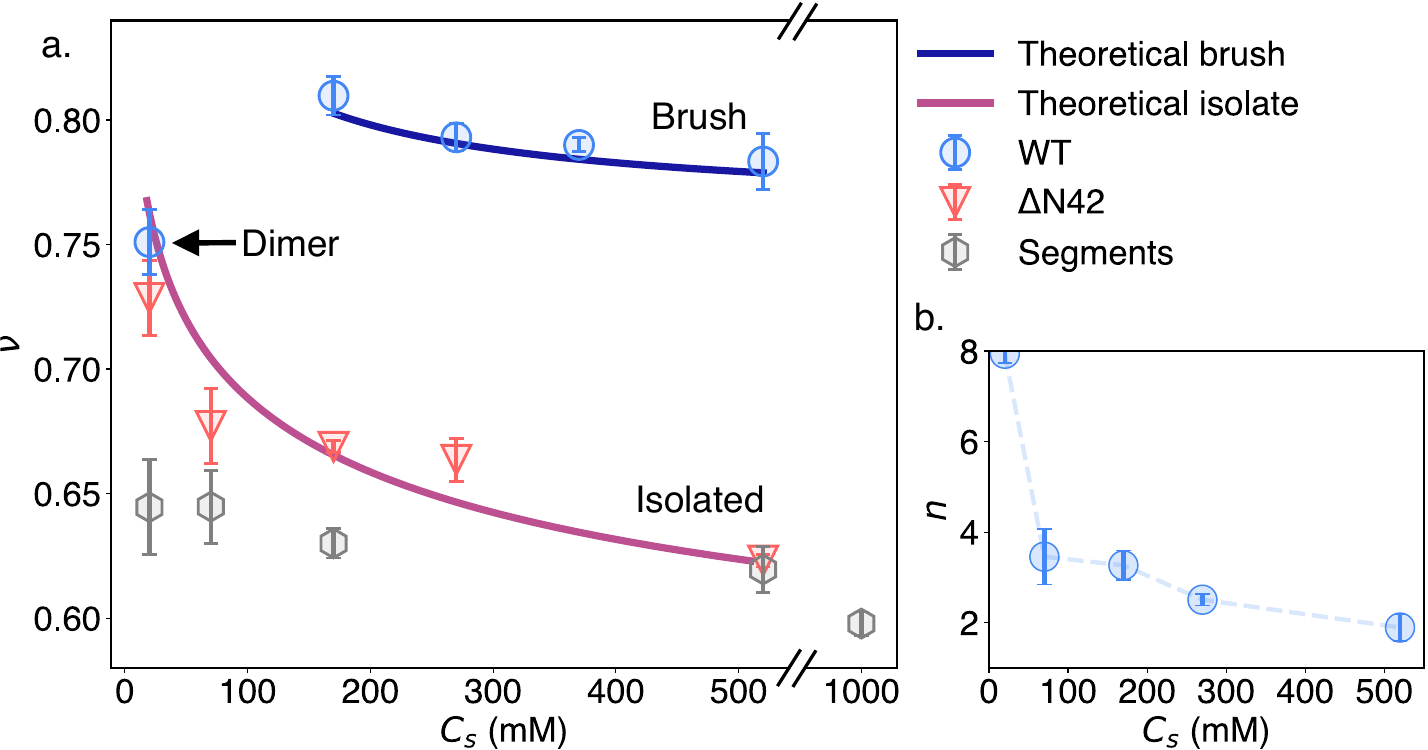}
\caption{Deduced structural parameters from the SAXS data fitting. \textbf{a.} Flory exponent ($\nu$) of WT tails and \dn variants showing extended disordered scaling. The red line refers to the theoretical brush model \cite{kumar2005polyelectrolyte}, and the blue line refers to the theoretical polyelectrolyte \cite{ha1992conformations}. \dn shows a decrease in the protein extension due to the decline in intermolecular electrostatic repulsion (see also Fig. \ref{fig:zimm}). WT shows an increase in the extension when shifting from a dimer to a trimer, followed by a slight decline with a further increase in salinity. In gray, average $\nu$ is obtained from measuring separate NFLt segments with an NCPR of -0.3 to -0.6 \cite{koren2023intramolecular}. \textbf{b.} The core (aggregated) peptide length per polypeptide as a function of salinity. At high salinity, each polypeptide aggregates via 2--3 amino acids that form the star--like polymer core. Both panels' values are the extrapolated ‘zero concentration’ parameters (supplementary Fig. S8). }
\label{fig:nu_n}
\end{figure*}

The repulsion between the grafted tail is further emphasized when comparing $h/2$ for WT to the equivalent \dn length-scale ($R_{\rm G}$), showing a significant extension for WT (Fig. \ref{fig:fitting}e). We notice that the WT tail's length ($h$) increases at low salt (during the transitions from a dimer to a trimer), followed by a steady mild decrease as the $C_s$, and following $Z$ increase. Similar compactness with increasing $C_s$ is shown for \dn and is expected for polyelectrolyte due to the reduction in electrostatic repulsion \cite{wang2019effects}. 
To better compare the statistical structure of two variants of disordered regions, we followed the polymeric scaling notation $\nu$ that quantifies the compactness of the chain. For \dnp, we extracted $\nu$ from Eqs. \ref{eq:1} and \ref{eq:2} and found a significant decrease in its value as 50 mM of NaCl is added to the 20 mM Tris buffer (Fig. \ref{fig:nu_n}a). The following monotonic decline is in line with polyelectrolytic models and electrostatic screening effects \cite{ha1992conformations}, shown in a solid red line in Fig. \ref{fig:nu_n}a. Interestingly, previous measurements of segments within the NFLt charged domain were shown to have similar $\nu$ values as in \dn. However, the same decline in salinity was not observed (Fig. \ref{fig:nu_n}a) \cite{koren2023intramolecular}. 
For the WT variant, the scaling factor ($\nu$) of the `star--like polymer' brushes is extracted from Eq. \ref{eq:2}. Here, we use $R_{\rm G}=h/2$, where $h$ is obtained from Eq. \ref{eq:3}. For $C_s = 20$ mM, we find \replaced{that }{a} $\nu$ is of similar scale as for \dn. This similarity can be attributed to the nature of the dimer, where the \replaced{intramolecular } {intermolecular} electrostatic interactions dominate the expansion of each of the two tails. As $C_s$ increases by $150$ mM, $\nu$ exhibits a considerable increase, presumably due to neighboring tail repulsion. Above $C_s =170$ mM, $\nu$ shows a weak decrease. We attribute this weak decline to the salt--brush regime of polyelectrolyte brushes \cite{kumar2005polyelectrolyte} shown in solid blue in Fig. \ref{fig:nu_n}a. In this regime, $h\propto C_s^{-1/3}$, and subsequently $\nu\propto -\frac{1}{3}log(C_s)$. 

We note that the cores of the star--like polymers are relatively small and that each polypeptide aggregates through only a few, most likely hydrophobic, amino acids. From the tabulated amino-acid partial volume, $\langle \phi_{aa}\rangle$ \cite{zamyatnin1972protein}, \replaced{we estimate the comprising amino acids as spheres of volume $\langle \phi_{aa}\rangle$. From here, the average number of amino acids per polypeptide inside the core is estimated by the number of spheres that can fit within the core volume, divided by the aggregation number: $n= V_{\rm core}/(\langle \phi_{aa}\rangle\cdot Z)$. }{we evaluated the average number of amino-acids per polypeptide inside the core $n= V_{\rm core}/(\langle \phi_{aa}\rangle\cdot Z)$.}
\added{Noticeably, our fit results with small $n$ values, ranging between $\sim 7-2$ residues on average within the aggregate ensemble and depending on the buffer salinity. Attempting to `fix' $n$ to a larger constant residue per tail number results in a poorer fitting (Fig. S9).}
In Fig. \ref{fig:nu_n}a, we indeed see that the most significant change occurs at the low salt regime, where $n$ drops from an average of 7 to 3 amino acids ($C_s = 20, 170$~mM, respectively). \deleted{We presume this}\added{Such }behavior \deleted{to be due to a salting--in effect, as} is known to occur within globular proteins \cite{okur2017beyond} and were recently alluded to impact IDPs \cite{wohl2021salt}. The following trend is a further decrease in $n$, albeit much weaker, which results in a final average $n$ of about two as the salinity reaches $C_s=520$~mM. 

\begin{figure*}%[hbt!]
\centering
\includegraphics[width=0.8\linewidth]{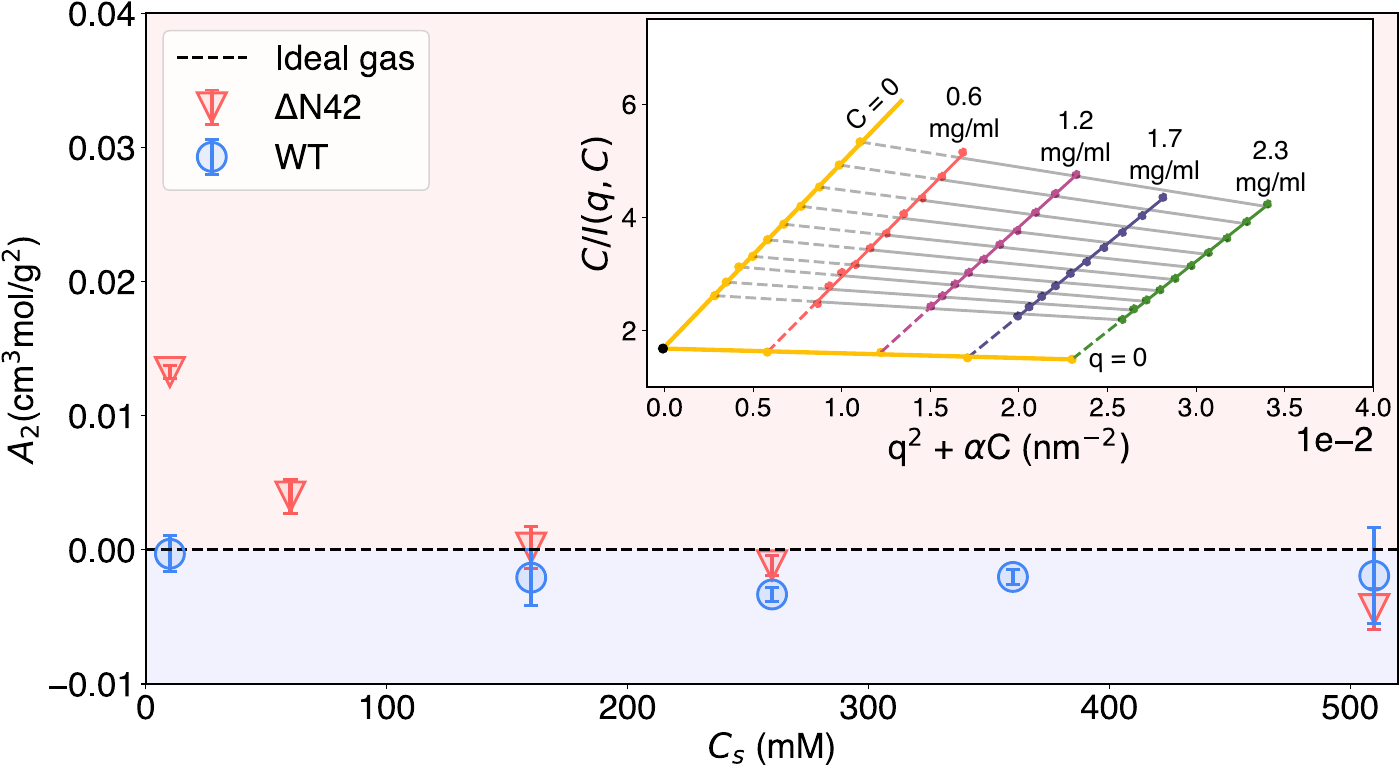}
\caption{The osmotic second virial coefficient $A_2$ as a function of the two variants' salinity ($C_s$). \dn intermolecular interactions transition from repulsive to attractive as $C_s$ increases. WT changes from a nearly neutral state of intermolecular interactions to attractive. \textbf{Inset}: A demonstration (WT variant, 20 mM Tris and 500 mM NaCl \textit{p}H 8.0) for the Zimm analysis used to extract $A_2$ from SAXS data measured at various protein concentrations ($C$). Values shown in the graph are in mg/ml units. The dashed lines show the extrapolation from the measured data (colored lines) to the fitted $q\rightarrow0$ and $C\rightarrow0$ yellow lines, where $\alpha=0.01$ is an arbitrary constant used in this analysis.}
\label{fig:zimm}
\end{figure*}

Last, in Fig. \ref{fig:zimm}, we quantify the intermolecular interactions by evaluating the second virial coefficient, $A_2$, using a Zimm analysis \cite{zimm1948scattering} (Table S5). Here, $A_2$ describes the deviation of the statistical ensemble from an ideal gas. In agreement with our previous data, we find that the inter--molecular interactions of \dn change from repulsive ($A_2>0$) to weakly attractive ($A_2 \leq 0$) as the salinity increases. In contrast, for WT, $A_2$ changes from a nearly neutral state of intermolecular interactions (\text{i.e.}, ideal gas regime) to mildly attractive ($A_2<0$). These findings are reflected in the dependency of the variant Flory coefficient $\nu$ in concentration. While at the lowest salinity, \dn is shown to expand as protein concentration is decreased, for higher salinities and for the WT measurements, $\nu$ remain primarily unchanged (Fig. S8a). 

Combining our results for both variants, we find an exemplary role of long--range electrostatic interactions tuning the statistical structure of IDPs. Without the uncharged N-terminal domain, the NFLt exhibited significant change as the electrostatic interactions were screened, causing them to condense further. In contrast, the presence of the uncharged domain incurred aggregation of the proteins, bringing the tails much closer to each other. The increase in proximity was reflected in a significant increase in the expansion compared to the truncated variant, which exhibited a much weaker contraction with salinity.
 
\section*{Discussion and Conclusions}

We investigated the effects of \replaced{sequence }{structural} heterogeneity on the interactions of NFLt, an IDP model system. For NFLt, the N-terminal region consisting of the first $\sim 50$ residues is hydrophobic and charge neutral, while the remaining chain is highly charged. We found that the sequence heterogeneity differentiates between the structures of the entire WT NFLt and a variant lacking the N-terminal domain. In particular, the WT variant self-assembles into star-like structures while the \dn one remains isolated in all measured cases. 

Since \dn can be attributed as a charged polymer, weakly attractive interactions take center stage as the electrostatic repulsion diminishes with charge screening (Fig. \ref{fig:zimm}). These interactions could be attributed to monomer-monomer attractions that arise from the sequence heterogeneity of the IDP, such as weak hydrophobic attraction from scattered hydropathic sites \cite{koren2023intramolecular,kornreich2016neurofilaments,beck2010gel,uversky1999natively,moglich2006end,pappu2000flory}.

For the WT variant, the intermolecular interactions started from a near-neutral state and transitioned to weakly attractive. However, as the WT measurements describe self-assembling complexes, the interpretation of these results differs from \dnp. As such, we interpret the intermolecular interactions as the `aggregation propensity,' the protein complex's growing ability. The aggregation propensity grows as the attractivity between the complex and the other polypeptides in the solution increases. This behavior can be observed when examining the responsiveness of the aggregation number $Z$ to protein concentration $C$ (Fig. S7). In the lowest measured screening, $Z$ dependency on protein concentration was minimal. As we increase the screening effects, this dependency becomes more substantial. This characterization is also found in folded proteins, where intermolecular interactions were shown to indicate aggregation propensity \cite{quigley2015second}. \added{The increased intermolecular attraction induced at increasing salinity is indicative of a salting--out phenomenon \cite{nandi1972aeffects,nandi1972beffects}, although further investigation at higher salinity is needed.} 

\added {The stability of the star-like polymer core should be evaluated by the participating residues per polypeptide ($n$). Indeed, while our fittings result with rather small $n$ values, the SAXS signal at low $q$ is dominated with aggregated structures under all salinity conditions. \replaced{Within the occurring hydrophobic interactions, the release of bound water molecules and ions from the polypeptides is likely to contribute to the core’s stability }{In addition to the hydrophobic bonds, the release of bound water molecules and ions from the polypeptides is likely to contribute to the core’s stability}. Such entropic based effects have been observed in similar processes such as protein flocculation \cite{dickinson2019strategies, blijdenstein2004depletion} and in temperature specific IDP binding modulation \cite{zosel2020depletion}.}

In our previous study \cite{koren2023intramolecular}, Flory exponents ($\nu$) of shorter segments from the same NFLt were measured independently and in the context of the whole NFLt using SAXS and time-resolved F{\"o}rster resonance energy transfer (trFRET). There, regardless of the peptide sequence, in the context of the entire NFLt, the segments' structural statistics were more expanded (\textit{i.e.,} with larger $\nu$ values) than when measured independently. Similarly, these short segments measured with SAXS have smaller $\nu$ values (\text{i.e.}, with a compacted statistical structure) than those of measured here for \dn in all salt conditions (Fig. \ref{fig:nu_n}a, grey symbols). 

The expansion of segments in the context of a longer chain corroborates that long-range contacts contribute to the overall disordered ensemble \cite{koren2023intramolecular}. Interestingly, at $C_s = 520$ mM salinity, we found similar $\nu$ values of the \dn and the previous short segment measurements, indicating a comparable expansion. We suggest that at higher salinities, the significance of electrostatic long-range contacts diminishes, aligning the expansion `scaling laws' regardless of the chain length. Importantly, comparisons between our \dn variant results (and not to the WT variant) to the previous segments' measurements are more suitable as the chains did not aggregate in those cases.

Compared to \dnp, WT exhibits a mild contraction in salt, resembling the behavior of the `salt--brush' regime observed in polyelectrolyte brushes, as demonstrated in Fig. \ref{fig:nu_n}. Similar salt--brush behavior was previously observed in Neurofilament high tail domain brushes grafted onto a substrate \cite{srinivasan2014stimuli}, and in a recent polyelectrolytic brush scaling theory \cite{zhulina2023cylindrical}. In the salt-brush regime, Pincus showed that brush mechanics resemble neutral brushes, determined by steric inter-chain interactions \cite{pincus1991colloid}. In this interpretation, the effective excluded volume per monomer enlarges and is proportional to $1/\kappa_{\rm s}^2$, where $\kappa_{\rm s}$ is the Debye length attributed to the added salt. Consequently, we suggest that the heightened charge screening in the WT solution allows steric interactions between brushes to play a more significant role in determining the brush ensemble. Additionally, we deduce that the increased prevalence of steric repulsion counteracts the attractive forces responsible for aggregation, thereby preventing brush collapse.

The NFLt contraction aligns with previous studies of native NFL hydrogel networks \cite{kornreich2016neurofilaments,beck2010gel}. At high osmotic pressure, the NFL network showed weak responsiveness to salinity higher than $C_s = 100$ mM, in agreement with theory \cite{zhulina2023cylindrical}. With the observed salt--brush behavior for WT, we suggest that weak salt response in NFL hydrogels coincides with the increase in steric repulsion shown for the star-like structures (Fig. \ref{fig:nu_n}a, blue line). 

Additionally, our measurements show that the hydrophobic N-terminal regime of the NFLt domain aggregates. This result is consistent with the findings of Morgan et al. \cite{morgan2020glassy}, where single-molecule pulling experiments were performed on WT NFLt, and slow aging effects were observed, likely due to collapse (and potential aggregation) of the neutral domain. Indeed, follow--up studies by Truong et al. \cite{truong2023pincus} used single-molecule stretching to show that added denaturant led to a swelling of the chain (increased $\nu$), demonstrating that the WT chain has hydrophobic aggregation that can be disrupted by the denaturant. These observations suggest that at higher salt, the loss of repulsion may lead to attractive hydrophobic interactions growing more prominent in the NFL network. However, the steric repulsion from the remaining NFL tail may shield such an unwanted effect. Nonetheless, such effects may grow more prominent as the native filament assembly is disrupted. 

In summary, we showed how the sequence composition of the NFLt IDP caused structural deviation from a disordered polyelectrolyte to a self--assembled star--like polymer brush. Together with the self--regulatory properties of the brushes, such behavior can be exploited to design structures that can resist specific environmental conditions. Additionally, our results showed possible implications on NFL aggregates that could shed light on the underlying correlations between the complex structure and the conditions driving it. While IDPs resemble polymers in many aspects, as we showed here, it is critical to assess their sequence to distinguish where and how to use the appropriate theoretical arguments to describe their statistical properties and structure.

\section*{Methods}
\textbf{Protein purification} Protein purification followed Koren et al. \cite{koren2023intramolecular}. Variant \dnp, included two cysteine residues at the C- and N terminals. After purification, \dn variants were first reduced by 20 mM 2-Mercaptoethanol. Next, 2-Mercaptoethanol was dialysed out with 1 L of 50 mM HEPES at pH 7.2. To block the cysteine sulfhydryl group, we reacted \dn variants with 2-Iodoacetamide at a molar ratio of 1:20. At the  reaction, the variants' concentrations were $\sim$2 mg/ml. The reaction solution was kept under dark and slow stirring for 5 hr and stopped by adding 50 mM 2-Mercaptoethanol followed by overnight dialysis against 1 L of 20 mM Tris at pH 8.0 with 0.1\% 2-Mercaptoethanol. Final purity was $>$95\% as determined by SDS-PAGE (Fig. S10). 

\textbf{SAXS measurement and analysis} Protein samples were dialyzed overnight in the appropriate solution and measured with a Nanodrop 2000 spectrophotometer (Thermo Scientific) for concentration determination. Buffers were prepared with 1 mM of TCEP to reduce radiation damage and 0.2\% of Sodium Azide to impair sample infection. The samples were prepared in a final concentration of 2 mg/ml, measured in a series of 4 dilutions. Preliminary measurements were measured at Tel-Aviv University with a Xenocs GeniX Low Divergence CuK$\alpha$ radiation source setup with scatterless slits \cite{li2008scatterless} and a Pilatus 300K detector. All samples were measured at three synchrotron facilities: beamline B21, Diamond Light Source, Didcot, UK \cite{cowieson2020beamline},  beamline P12, EMBL, DESY, Hamburg, Germany \cite{blanchet2015versatile}, and beamline BM 29 ESRF, Grenoble, France \cite{pernot2013upgraded}. Measurements at ESRF were done using a robotic sample changer \cite{round2015biosaxs}.  

Integrated SAXS data was obtained from the beamline pipeline and 2D integration using the "pyFAI" Python library \cite{kieffer2020new}. Extended Guinier analyses for the \dn variant were done with the "curve\_fit" function from the "Scipy" Python library \cite{virtanen2020scipy}. To extract $R_g$ and $\nu$, extended Guinier analysis was conducted for $0.7 < qR_g < 2$. Error calculation was done from the covariance of the fitting.

Model fittings for the WT variant were done using the "lmfit" Python library \cite{newville2016lmfit} using the model described in \cite{pedersen2002scattering,pedersen2000form}. Due to the complexity of the model, cylindrical core fittings were done by binning the data in 100 logarithmic bins to reduce computation time. Within the same model, core parametres (cylinder radius $R$ and cylinder length $L$) were set constant, to offset fitting errors. Initial values of $R$ and $L$ were calculated with the highest measured concentration. Physical boundary conditions were imposed on the fitting, and scattering length (SL) values were set to be unchanged by the fitting process. SL values of both the core and the tail domains were determined by tabulated values of amino acid SLD in 100\% H$_2$O \cite{jacrot1976study} (Table S3). Fitting parameter error evaluation was done by finding the covariant of the returning fitting parameters. Error calculation of the volume was done using: 
$\frac{dV}{V} = \sqrt{3 \left (\frac{dR}{R}\right)^2}$. In addition,
$\nu$ values of WT were found by a recursive search of the corresponding tail height $h / 2$ over Eq. \ref{eq:2}. Errors of $\nu$ were then found by assuming a simple case of $R_g = b N^{\nu}$, from which:
$d\nu \sim \frac{\ln{(1 + dR/R)}}{\ln{N}} \sim \ln{(N)}^{-1}\frac{dR}{R}$ 

\textbf{Zimm analysis}
Zimm analysis was performed as described in \cite{zimm1948scattering}. Data normalization was done by first determining $I_0$ by fitting a linear curve over the Guinier plot ($\ln{I(q)}$ vs $q^2$). Normalized $1/I(q)$ linear fitting was done starting with the earliest possible data point until a deviation from the linear behavior occurs. Data points were then binned for visual clarity without impacting the result. 

\textbf{Brush model fitting}
Brush height model as described in \cite{kumar2005polyelectrolyte} was fitted with a prefactor $c = 0.33$ to match data. Resulting heights were converted to $\nu$ by $h = bN^\nu$ where $b=0.38$ nm and $N=146$. To accommodate for the change in grafting density, a linear curve was fitted to the grafting density's change in salinity and was used to obtain a continuous plot.  

\textbf{Polyelectrolye fitting}
The fitting model was used as described in \cite{ha1992conformations} with a pre-factor $c=1.24$ to match data.

\section*{Acknowledgments}
R.B. and O.A.S. dedicate this article to Fyl Pincus, for his continuous leadership and friendship over the years. His past works on charged polymer brushes, and polymers' scaling laws, inspired much research in the field, including this work. 
The synchrotron SAXS data were collected at beamline P12, operated by EMBL Hamburg at the PETRA III storage ring (DESY, Hamburg, Germany), at beamline B21, operated by Diamond Light Source (Didcot, UK), and at beamline BM29, operated by ESRF (Grenoble, France). We would like to thank Cy M. Jefferies (DESY), Katsuaki Inoue (DLS), and Mark Tully (ESRF)  for their assistance in using the beamlines. 
This work has been supported by the NSF (MCB-2113302), the  NSF–BSF program (2020787), the Israel Science Foundation (1454/20), and by iNEXT-Discovery (15410), funded by the Horizon 2020 program of the European Commission. We also acknowledge the fruitful discussion and help from Yacov Kantor, Uri Raviv, and Sagi Meir.

\subsection*{Statements and Declarations}
\textbf{Conflicting interests} The authors claim no conflicting interests. \\
\textbf{Data availability} The raw SAXS data is available in the Small--Angle Scattering Biological Data Bank (SASBDB) at: \url{https://www.sasbdb.org/project/2190/}. \\
\textbf{Author contribution} M.K., G.K., and R.B. designed the project. M.K. conducted experiments and analysis with G.K.'s and O.A.S.'s assistance. M.K., G.K., R.B., and O.A.S. wrote the paper.

\bibliography{sn-bibliography}

\clearpage
\onecolumn
\section*{Supplementary Information}
\textbf{\large From isolated polyelectrolyte to star-like assemblies:\\ the role of sequence heterogeneity on the statistical structure of the intrinsically disordered Neurofilament-low tail domain} \\
\text{\small Mathar Kravikass, Gil Koren, Omar Saleh, Roy Beck} \\
\text{\small Corresponding E-mail:roy@tauex.tau.ac.il} \\\\
\\
\\
\textbf{Contents:}
\\
Table S1-S5, \\
Figure S1-S10
\\
\\
\\

%\pagebreak

\setcounter{figure}{0}
\setcounter{table}{0}
\renewcommand{\thefigure}{S\arabic{figure}}
\renewcommand{\theHfigure}{S\arabic{figure}}
\renewcommand{\thetable}{S\arabic{table}}
\renewcommand{\theHtable}{S\arabic{table}}

\begin{figure*}[ht!]
\centering
\includegraphics[width=1\linewidth]{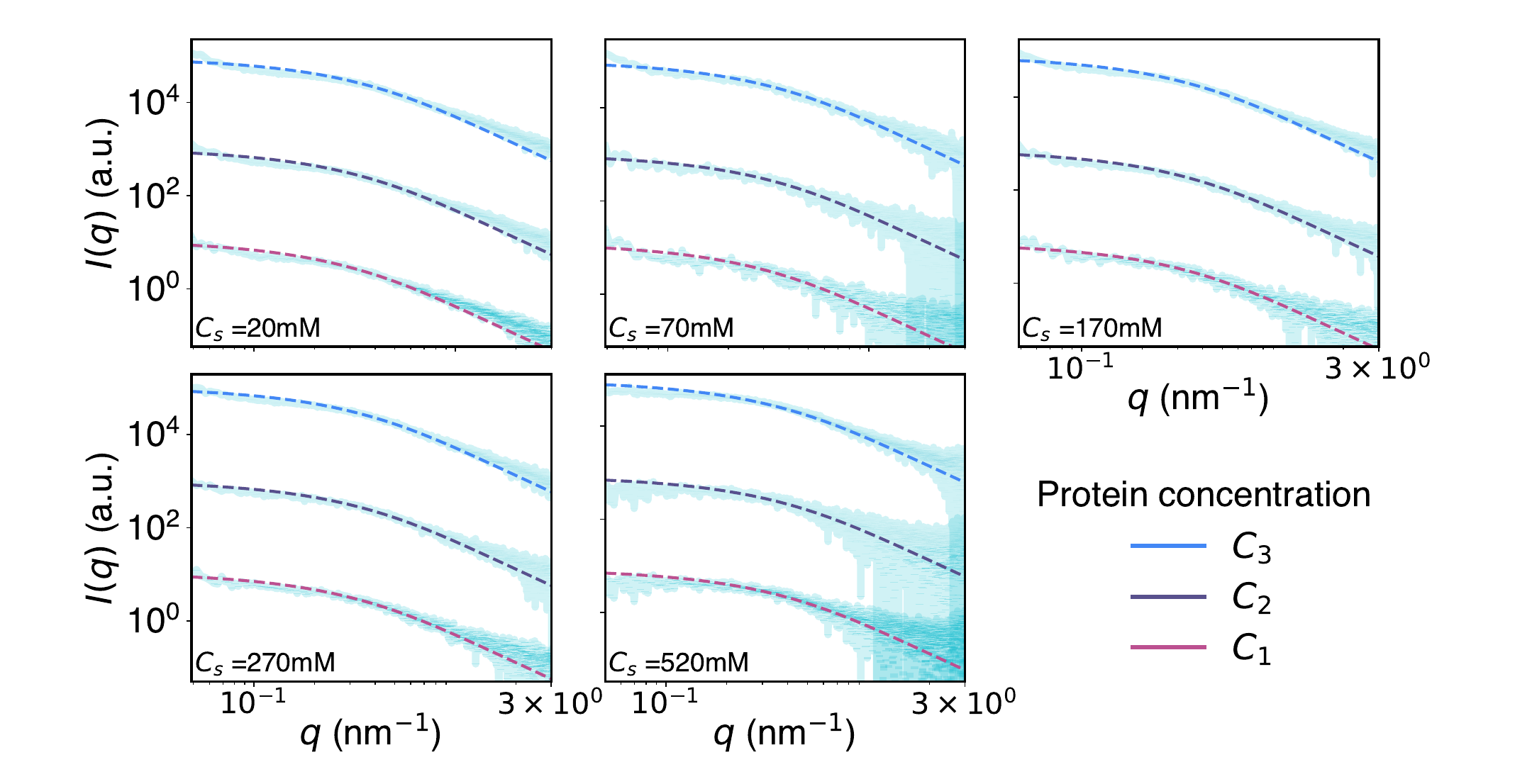}
\caption{\dn SAXS measurements Gaussian form factor fitting for all salinity concentrations $C_s$. $R_G$ used for the Gaussian form factor is as obtained by the Extended Guinier analysis (see table \ref{tab_S:dn}).}
\label{fig:dn_alL_gauss_fittings}
\end{figure*}

\begin{figure*}[ht!]
\centering
\includegraphics[width=1\linewidth]{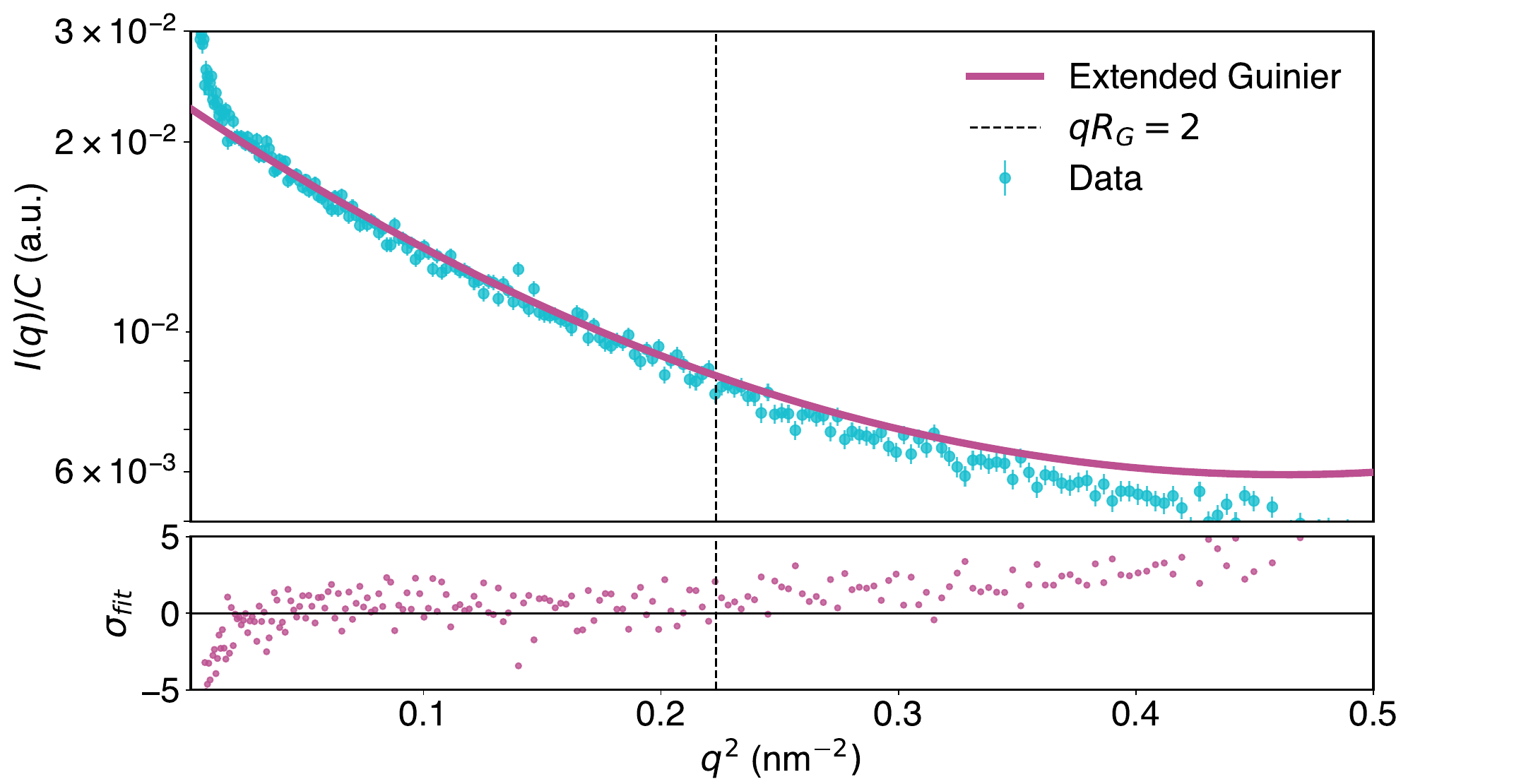}
\caption{\dn SAXS measurement with corresponding extended Guinier curve. Bottom: Deviation from fit $\sigma_{fit} = (Y_{data}-Y_{fit})/\sigma_{data}$ Dashed line represents the maximum analysis point $qR_G = 2$ from which deviation starts. Displayed data: $20$mM Tris pH$8.0$ in $1.1$mg/ml.}
\label{fig:fig_dn_extended_fit}
\end{figure*}

\begin{figure*}[ht!]
\centering
\includegraphics[width=1\linewidth]{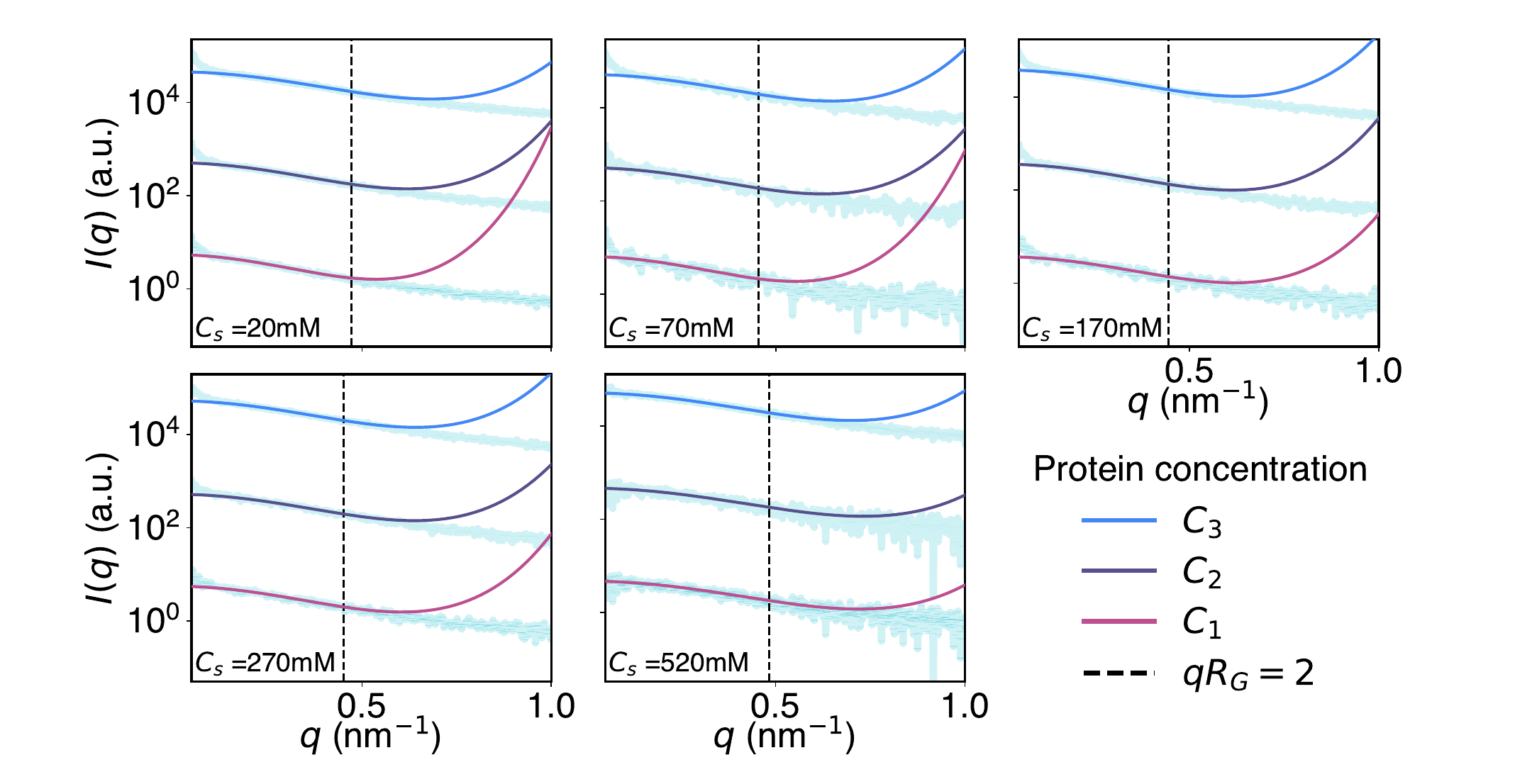}
\caption{\dn SAXS measurements with corresponding extended Guinier curves. Dashed lines represent the maximum analysis point $qR_G = 2$ from which deviation starts. Protein concentrations were offset for clarity, with the lowest (blue) being of the highest concentration.}
\label{fig:nu_all_extended}
\end{figure*}

\begin{table*}[!ht]
 %   \resizebox{\textwidth}{!}
 {
    \centering
    \begin{tabular}{|l|l|l|l|l|}
    \hline
        $C_s$ (mM) & C (mg/ml) & $R_G$ (nm) & $\nu$ & $I_0$ (cm$^{-1}$) \\ \hline
        20 & 1.1 & 4.23 $\pm $ 0.05 & 0.642 $\pm $ 0.002 & 0.0228 \\ \hline
        20 & 0.8 & 4.56 $\pm $ 0.07 & 0.660 $\pm $ 0.66 & 0.0254 \\ \hline
        20 & 0.6 & 5.11 $\pm $ 0.13 & 0.689 $\pm $ 0.007 & 0.027 \\ \hline
        70 & 1 & 4.41 $\pm $ 0.12 & 0.652 $\pm $ 0.007 & 0.0259 \\ \hline
        70 & 0.5 & 4.53 $\pm $ 0.22 & 0.659 $\pm $ 0.012 & 0.0257 \\ \hline
        70 & 0.3 & 4.99 $\pm $ 0.46 & 0.683 $\pm $ 0.023 & 0.032 \\ \hline
        170 & 1.5 & 4.51 $\pm $ 0.05 & 0.657 $\pm $ 0.003 & 0.19 \\ \hline
        170 & 0.8 & 4.59 $\pm $ 0.11 & 0.662 $\pm $ 0.006 & 0.18 \\ \hline
        170 & 0.3 & 4.56 $\pm $ 0.27 & 0.661 $\pm $ 0.015 & 0.18 \\ \hline
        270 & 1 & 4.43 $\pm $  0.06 & 0.653 $\pm $ 0.003 & 0.026 \\ \hline
        270 & 0.5 & 4.45 $\pm $ 0.1 & 0.654 $\pm $ 0.006 & 0.026 \\ \hline
        270 & 0.3 & 4.64 $\pm $ 0.16 & 0.664 $\pm $ 0.009 & 0.028 \\ \hline
        520 & 1.5 & 4.14 $\pm $ 0.02 & 0.636 $\pm $ 0.001 & 0.026 \\ \hline
        520 & 0.78 & 4.00 $\pm $ 0.08 & 0.628 $\pm $ 0.005 & 0.023 \\ \hline
        520 & 0.38 & 4.05 $\pm $ 0.35 & 0.630 $\pm $ 0.02 & 0.023 \\ \hline
    \end{tabular}
    }
    \caption {\textbf {\dn Extended guinier analyis data}. Analysis parametres (radius of gyration $R_G$, scaling exponent $\nu$, and scattering intensity at $q=0$ ($I_0$)) obtained for different salt concentrations ($C_s$) and protein concentrations ($C$).}
    \label{tab_S:dn}
\end{table*}

\begin{table*}[!ht]
    %\resizebox{\textwidth}{!}
    {%
    \centering
    \centering
    \begin{tabular}{|l|l|l|}
    \hline
        $C_s$ (mM) & $R_G$ (nm) & $\nu$ \\ \hline
        20 & 5.76 $\pm $ 0.31 & 0.729 $\pm $ 0.015 \\ \hline
        70 & 4.84 $\pm $ 0.27 & 0.677 $\pm $ 0.015 \\ \hline
        170 & 4.71 $\pm $  0.04 & 0.669 $\pm $ 0.003 \\ \hline
        270 & 4.61 $\pm $ 0.16 & 0.663 $\pm $ 0.009 \\ \hline
        520 & 3.88 $\pm $ 0.04 & 0.620 $\pm $ 0.003 \\ \hline
    \end{tabular}
    }
    \caption {\textbf{Zero concentration extended Guinier analysis data}. Analysis parameters (radius of gyration $R_G$ and scaling exponent $\nu$) were extrapolated to zero protein concentration at various salt concentrations ($C_s$).}
    \label{tab_S:dn_zero}
\end{table*}

\begin{figure*}[hbt!]
\centering
\includegraphics[width=1\linewidth]{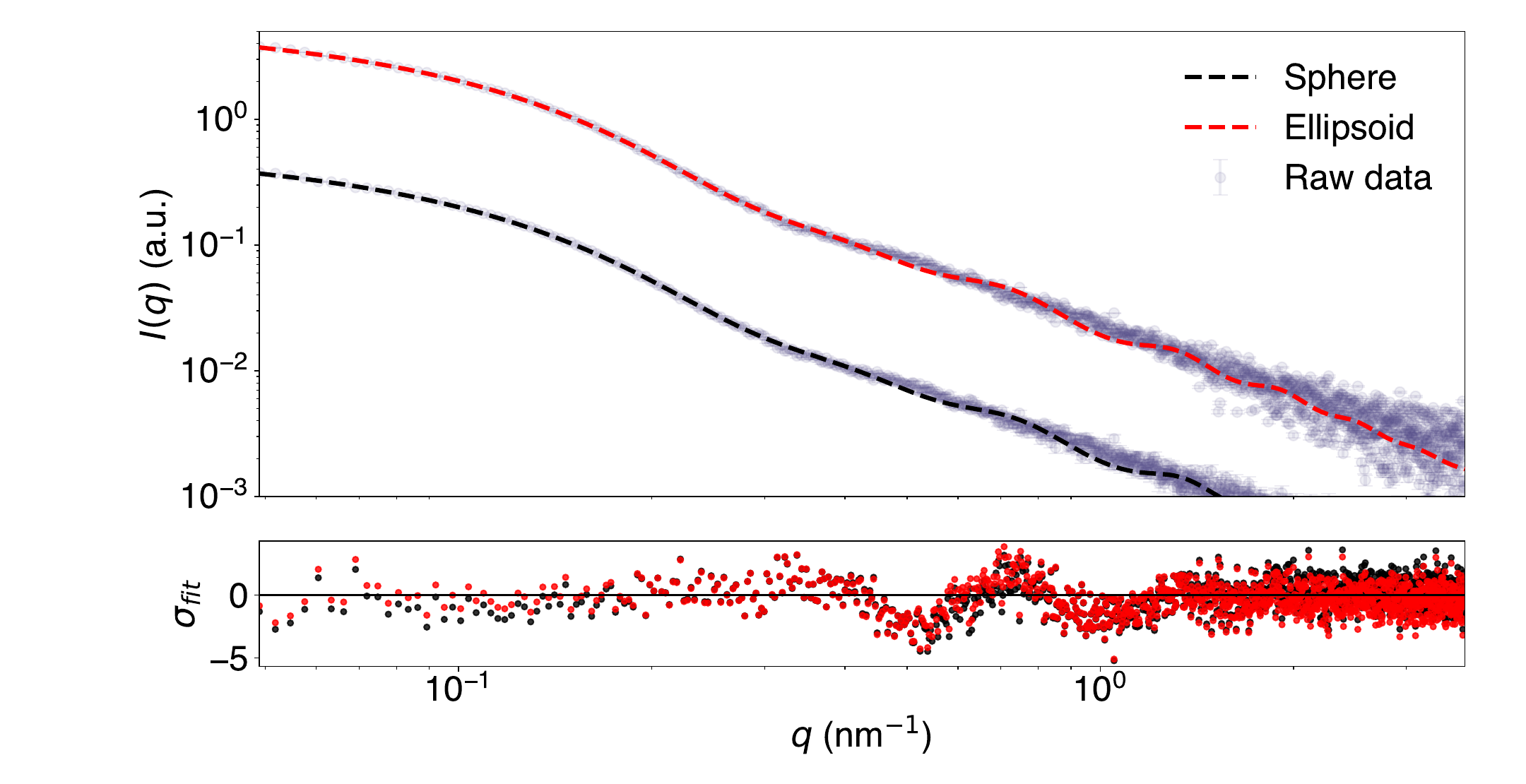}
\caption{SAXS measurements of WT and its fitting to different form factors. Both form factors are of the same model but use a different core: Spherical or Ellipsoidal. Spherical core fitting yields a core radius of $R = 0.66 \pm  0.016$ nm, and the ellipsoidal core yields a core radius of $R = 1.335 \pm  0.23$ nm and a secondary radius of $\epsilon R$ where $\epsilon = 0.153 \pm  0.08$. Both fittings yield close values of aggregation number Z ($3.046 \pm  0.04$ for spherical and $3.562 \pm  0.07$ for ellipsoidal) and tail height $h/2$ ($9.838 \pm  0.04$ nm for spherical and $9.584 \pm  0.12$ for ellipsoidal). Below: Fitting error $\sigma_{fit} = (Y_{fit}-Y_{data})/\sigma_{data}$. Both curves show similar error profiles. The spherical model proved best to describe the model due to its simplicity. Displayed data: WT in $20$ mM Tris pH=8.0, and $170$ mM NaCl at a concentration of $1.3$ mg/ml.  }
\label{fig:different_cores}
\end{figure*}

\begin{figure*}[hbt!]
\centering
\includegraphics[width=1\linewidth]{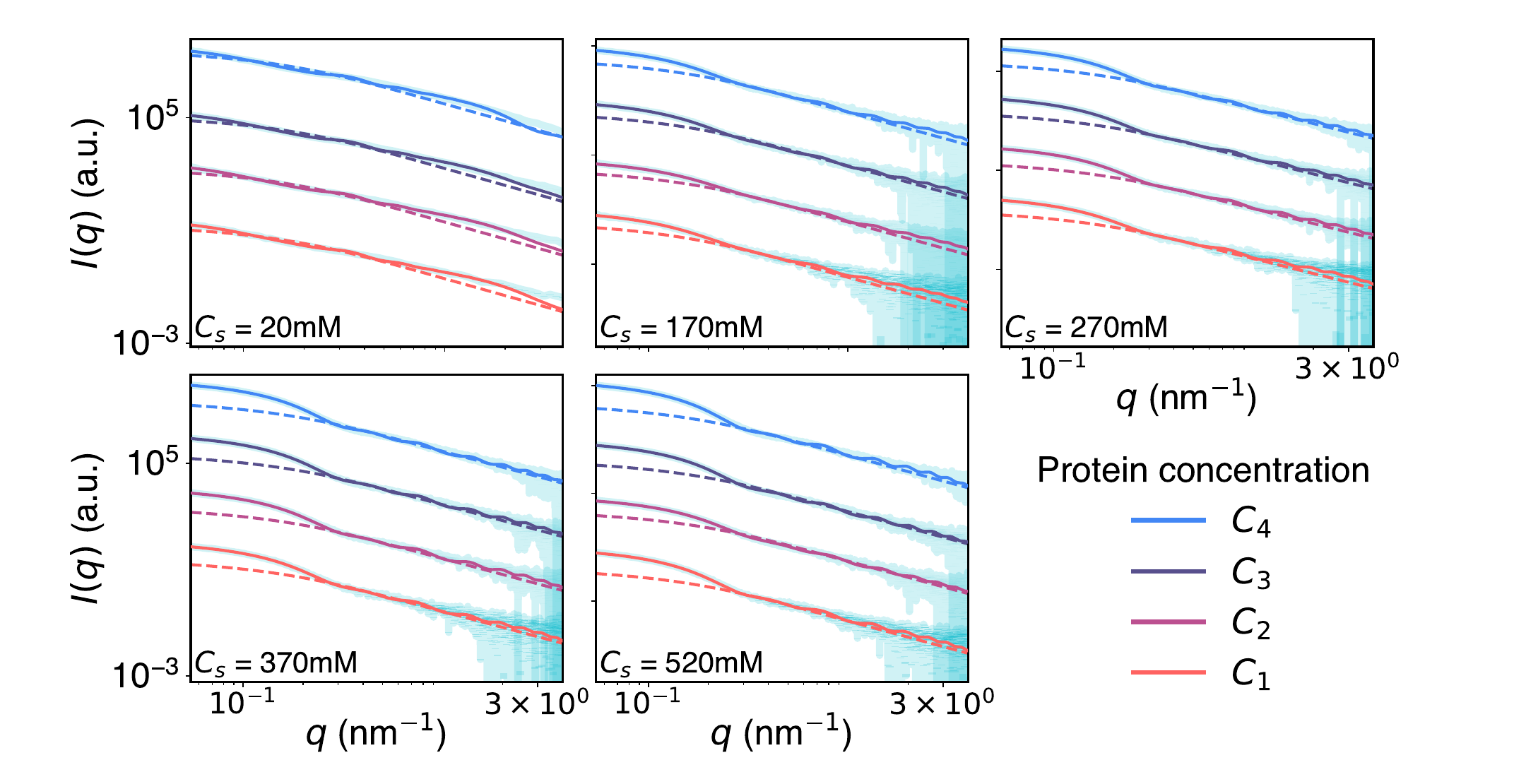}
\caption{SAXS measurements and spherical form factor fitting for all salinity concentrations ($C_s$). The $C_s = 20$ mM data fit is to a cylindrical core. Dashed lines represent the Gaussian form factor of the structure tails. Protein concentrations were offset for clarity, with the lowest (blue) being the highest concentration.}
\label{fig:different_fittings}
\end{figure*}

\begin{figure*}[ht!]
\centering
\includegraphics[width=1\linewidth]{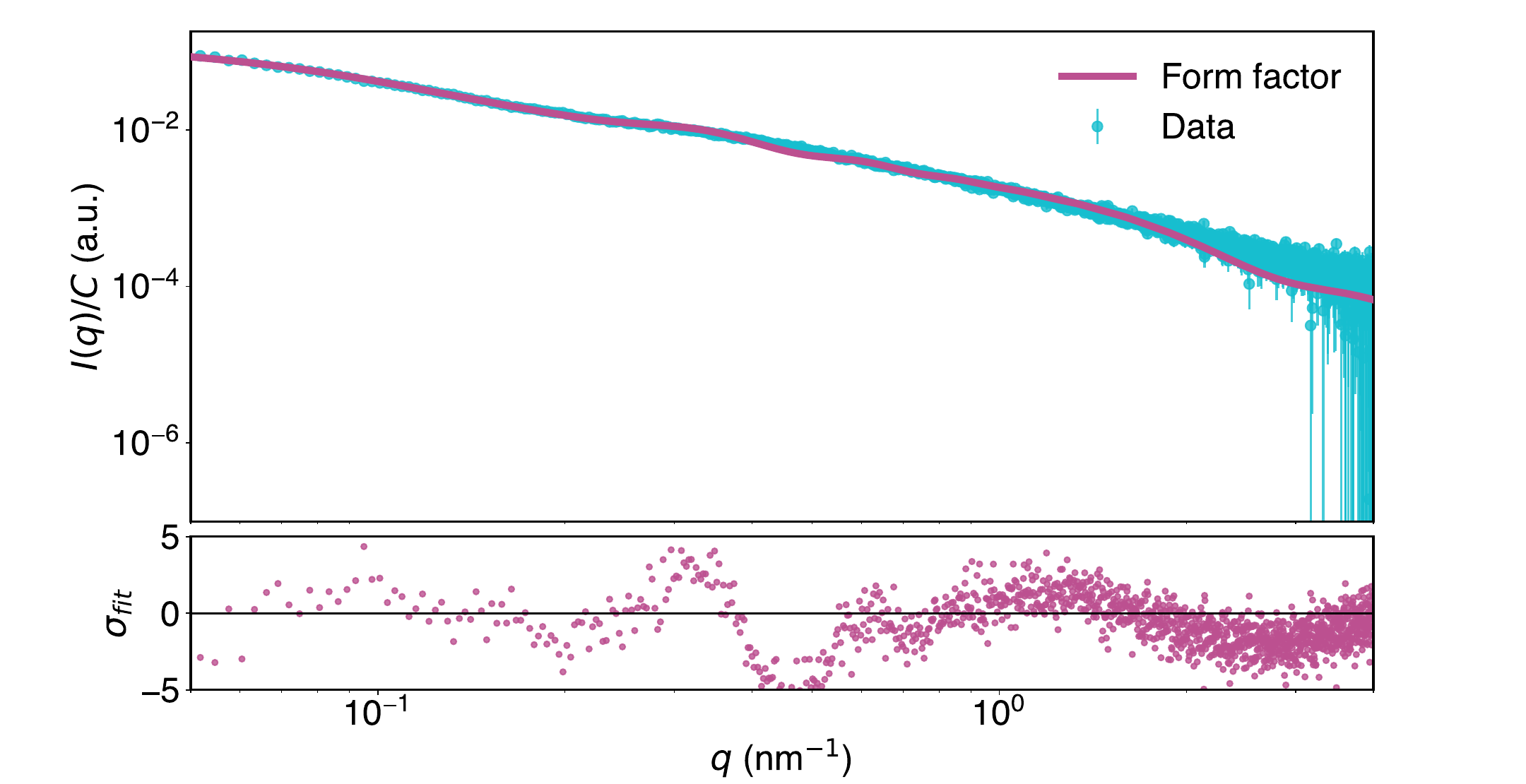}
\caption{WT SAXS measurement with cylindrical fitting. Measurements at the highest concentration of $C=2.68$ mg/ml, in a $20$ mM Tris buffer at pH=$8.0$. To alleviate fitting inconsistencies, consequent fittings of measurements with lower protein concentrations in the same buffer were done using the obtained core parameters: Core radius $R = 0.89\pm0.03$ nm, and core length $L = 1.19\pm0.09$ nm.}
\label{fig:fig_cylinder_fit}
\end{figure*}

\begin{table*}[!ht]
    \resizebox{\textwidth}{!}
   {
    \centering
    \begin{tabular}{|l|l|l|l|l|l|l|l|l|l|l|}
    \hline
     $C_s$& $C$ & $h/2$ & $\nu$ & $Z$ & $n$ & $R$ & $L$ & $V$ & $\beta_t$ & $\beta_c$ \\(mM) &(mg/ml)&(nm)& & & & (nm)& (nm) & (nm$^3$) & (10$^3$nm) & (10$^3$nm)\\  \hline
        20 & 2.68 & 9.16\textpm 0.15 & 0.786\textpm 0.0033 & 1.60\textpm 0.03 & 10.11\textpm 0.711 & 0.89\textpm 0.028 & 1.19\textpm 0.09 & 2.26\textpm 0.027 & 0.227 & 3.826 \\ \hline
        20 & 1.8 & 8.20\textpm 0.13 & 0.758\textpm 0.0032 & 1.83\textpm 0.04 & 8.52\textpm 0.059 & 0.89 & 1.19 & 2.26 & 0.195 & 3.558 \\ \hline
        20 & 1 & 8.57\textpm 0.17 & 0.768\textpm 0.0041 & 1.87\textpm 0.05 & 8.12\textpm 0.060 & 0.89 & 1.19 & 2.26 & 0.195 & 3.558 \\ \hline
        20 & 0.5 & 8.27\textpm 0.16 & 0.759\textpm 0.0040 & 1.91\textpm 0.05 & 7.78\textpm 0.057 & 0.89 & 1.19 & 2.26 & 0.195 & 3.558 \\ \hline
        170 & 1.3 & 9.96\textpm 0.03 & 0.796\textpm 0.0006 & 3.34\textpm 0.02 & 2.52\textpm 0.009 & 0.66\textpm 0.005 & X & 1.18\textpm 0.007 & 0.039 & 4.014 \\ \hline
        170 & 0.73 & 10.11\textpm 0.05 & 0.799\textpm 0.0009 & 3.27\textpm 0.03 & 2.32\textpm 0.014 & 0.63\textpm 0.007 & X & 1.06\textpm 0.010 & 0.039 & 4.014 \\ \hline
        170 & 0.57 & 10.37\textpm 0.08 & 0.806\textpm 0.0015 & 2.13\textpm 0.02 & 3.10\textpm 0.037 & 0.60\textpm 0.006 & X & 0.93\textpm 0.008 & 0.074 & 3.98 \\ \hline
        170 & 0.24 & 10.78\textpm 0.10 & 0.815\textpm 0.0019 & 2.71\textpm 0.05 & 3.23\textpm 0.025 & 0.66\textpm 0.011 & X & 1.23\textpm 0.017 & 0.074 & 3.98 \\ \hline
        270 & 2 & 9.40\textpm 0.02 & 0.781\textpm 0.0003 & 5.37\textpm 0.02 & 1.89\textpm 0.007 & 0.70\textpm 0.004 & X & 1.42\textpm 0.008 & 0.018 & 4.03 \\ \hline
        270 & 1.5 & 9.43\textpm 0.02 & 0.782\textpm 0.0004 & 5.02\textpm 0.02 & 1.93\textpm 0.009 & 0.69\textpm 0.005 & X & 1.36\textpm 0.009 & 0.018 & 4.03 \\ \hline
        270 & 0.69 & 9.85\textpm 0.04 & 0.793\textpm 0.0009 & 4.05\textpm 0.04 & 2.70\textpm 0.016 & 0.71\textpm 0.009 & X & 1.53\textpm 0.016 & 0.039 & 4.014 \\ \hline
        370 & 2.3 & 9.36\textpm 0.01 & 0.780\textpm 0.0003 & 6.68\textpm 0.03 & 1.48\textpm 0.008 & 0.69\textpm 0.005 & X & 1.38\textpm 0.009 & 0.018 & 4.036 \\ \hline
        370 & 1.5 & 9.44\textpm 0.02 & 0.782\textpm 0.0003 & 6.43\textpm 0.03 & 1.65\textpm 0.009 & 0.71\textpm 0.006 & X & 1.49\textpm 0.010 & 0.018 & 4.036 \\ \hline
        370 & 0.96 & 9.53\textpm 0.02 & 0.785\textpm 0.0004 & 5.85\textpm 0.03 & 2.08\textpm 0.010 & 0.74\textpm 0.006 & X & 1.70\textpm 0.012 & 0.039 & 4.014 \\ \hline
        370 & 0.6 & 9.81\textpm 0.03 & 0.792\textpm 0.0007 & 5.33\textpm 0.05 & 2.13\textpm 0.017 & 0.72\textpm 0.010 & X & 1.59\textpm 0.019 & 0.039 & 4.014 \\ \hline
        520 & 2.5 & 9.57\textpm 0.01 & 0.786\textpm 0.0003 & 8.15\textpm 0.04 & 1.82\textpm 0.008 & 0.79\textpm 0.005 & X & 2.07\textpm 0.012 & 0.018 & 4.036 \\ \hline
        520 & 1.19 & 9.28\textpm 0.02 & 0.778\textpm 0.0004 & 6.66\textpm 0.04 & 1.57\textpm 0.010 & 0.70\textpm 0.006 & X & 1.47\textpm 0.012 & 0.018 & 4.036 \\ \hline
    
        520 & 0.45 & 9.82\textpm 0.03 & 0.792\textpm 0.0005 & 6.36\textpm 0.06 & 1.94\textpm 0.016 & 0.74\textpm 0.010 & X & 1.72\textpm 0.020 & 0.039 & 4.014 \\ \hline
    \end{tabular}
    }
    \caption {\textbf{WT spherical and cylindrical fitting analysis data}. Analysis parameters (brush height ($h$), scaling exponent ($\nu$), aggregation number ($Z$), core peptide length ($n$), core radius ($R$), cylindrical core length ($L$), core volume ($V$), tail scattering length ($\beta_t$) and core scattering length ($\beta_c$)) obtained for different salt concentrations ($C_s$) and protein concentrations ($C$). Cylinder length $L$ values are only relevant to $C_s = 20$mM where a cylindrical core fit was used. For the cylindrical core, the same values of $L$ and $R$ were used for all concentrations to alleviate fitting errors (see Methods).}
    \label{tab_S:wt}
\end{table*}

\begin{table*}[!ht]
    \resizebox{\textwidth}{!}
    {%
    \centering
    \begin{tabular}{|l|l|l|l|l|l|l|l|}
    \hline
        $C_s$&$h/2$&$\nu$ & $Z$ & $n$ & $R$  & $L$ & $V$ \\ (mM)&(nm)& & & &(nm)&(nm)&(nm$^3$)\\ \hline
        20 & 8.01\textpm 0.46 & 0.751\textpm 0.013 & 2.03\textpm0.08 & 7.08\textpm 0.39 & 0.89 & 1.19 & 2.26 \\ \hline
        170 & 10.60\textpm 0.33 & 0.811\textpm 0.008 & 2.83\textpm 0.31 & 3.17\textpm 0.55 & 0.64\textpm 0.03 & X & 1.10\textpm 0.16 \\ \hline
        270 & 9.84\textpm 0.21 & 0.793\textpm 0.006 & 3.52\textpm 0.22 & 3.06\textpm 0.30 & 0.70\textpm 0.03 & X & 1.49\textpm 0.20 \\ \hline
        370 & 9.73\textpm 0.12 & 0.790\textpm 0.003 & 5.19\textpm 0.29 & 2.39\textpm 0.12 & 0.76\textpm 0.02 & X & 1.67\textpm 0.01 \\ \hline
        520 & 9.47\textpm 0.44 & 0.783\textpm 0.011 & 5.67\textpm 0.35 & 1.82\textpm 0.29 & 0.68\textpm 0.06 & X & 1.28\textpm 0.38 \\ \hline
    \end{tabular}
    }
    \caption {\textbf{Zero concentration WT spherical and cylindrical fitting analysis data}. Analysis parametres (brush height ($h$), scaling exponent ($\nu$), aggregation number ($Z$), core peptide length ($n$), core radius ($R$), cylindrical core length ($L$) and core volume ($V$)) were extrapolated to zero protein concentration at various salt concentrations ($C_s$). Cylinder length $L$ values are only relevant to $C_s = 20$mM where a cylindrical core was used.}
    \label{tab_S:wt_zero}
\end{table*}

\begin{figure*}[hbt!]
\centering
\includegraphics[width=1\linewidth]{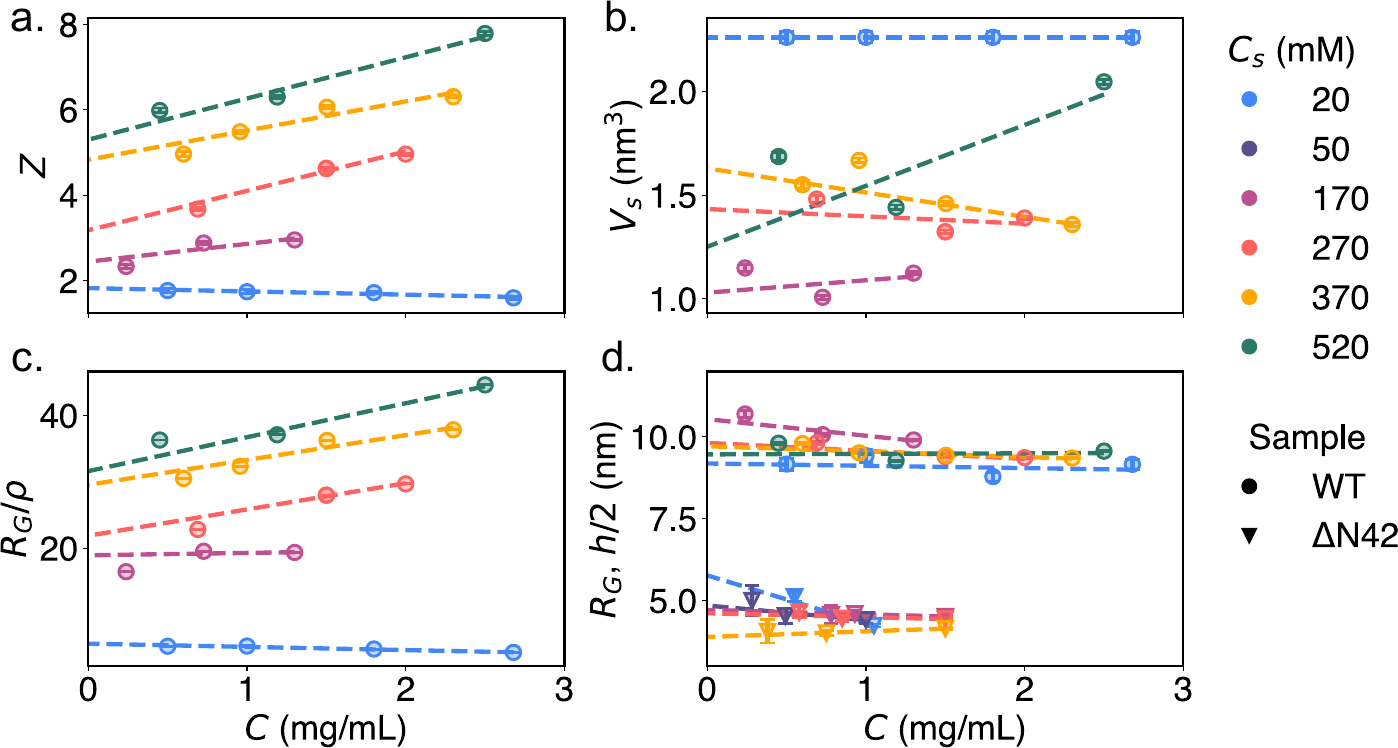}
\caption{Structural parameters for WT (circles) and \dn (triangles) variants extracted from fitting the SAXS data. Dashed lines demonstrate the linear fitting of the data used to obtained the zero concentration extrapolations. \textbf{a.} Aggregation number ($Z$) dependency on protein concentration ($C$) increases with increasing salt. \textbf{b.} Core volume $V_s$ against protein concentration ($C$). In $C_s =20$ mM, the $V_s$ values are constant due to fitting constraints (see Methods).  \textbf{c.} In all cases, the tail heights ($h$) are much larger than the corresponding grafting length ($\rho$), indicative of a brush regime. \textbf{d.} The structurally intrinsically disordered \dn variant compacts with higher $C_s$ values and remains more compacted from the projected tails for the WT variant. For the \dn variant $R_G$ drastically changes as a function of the protein concentration ($C$).}
\label{fig:sup_fitting}
\end{figure*}

\begin{table*}[!ht]{
    \centering
    \begin{tabular}{|l|l|l|}
    \hline
        $C_s$ & $A_2^{WT}$ & $A_2^{\Delta N42}$\\ 
        (mM) & (cm$^3$mol/g$^2\times 10^3$) & (cm$^3$mol/g$^2\times 10^3$)  \\
        
        \hline
        20 & -0.295\textpm1.346 & 13.264\textpm0.466 \\ \hline
        70 & X & 3.978\textpm1.248 \\ \hline
        170 & -2.072\textpm2.091 & 0.169\textpm1.544 \\ \hline
        270 & -3.328\textpm0.508 & -1.152\textpm0.756 \\ \hline
        370 & -2.020\textpm0.563 & X \\ \hline
        520 & -1.933\textpm3.582 & -4.417\textpm1.514 \\ \hline
    \end{tabular}
   }
    \caption{Second virial coefficient $A_2$ values for both variants in salt concentration $C_s$.}
    \label{tab_S:A2_params_}
\end{table*}

\begin{figure*}[hbt!]
\centering
\includegraphics[width=1\linewidth]{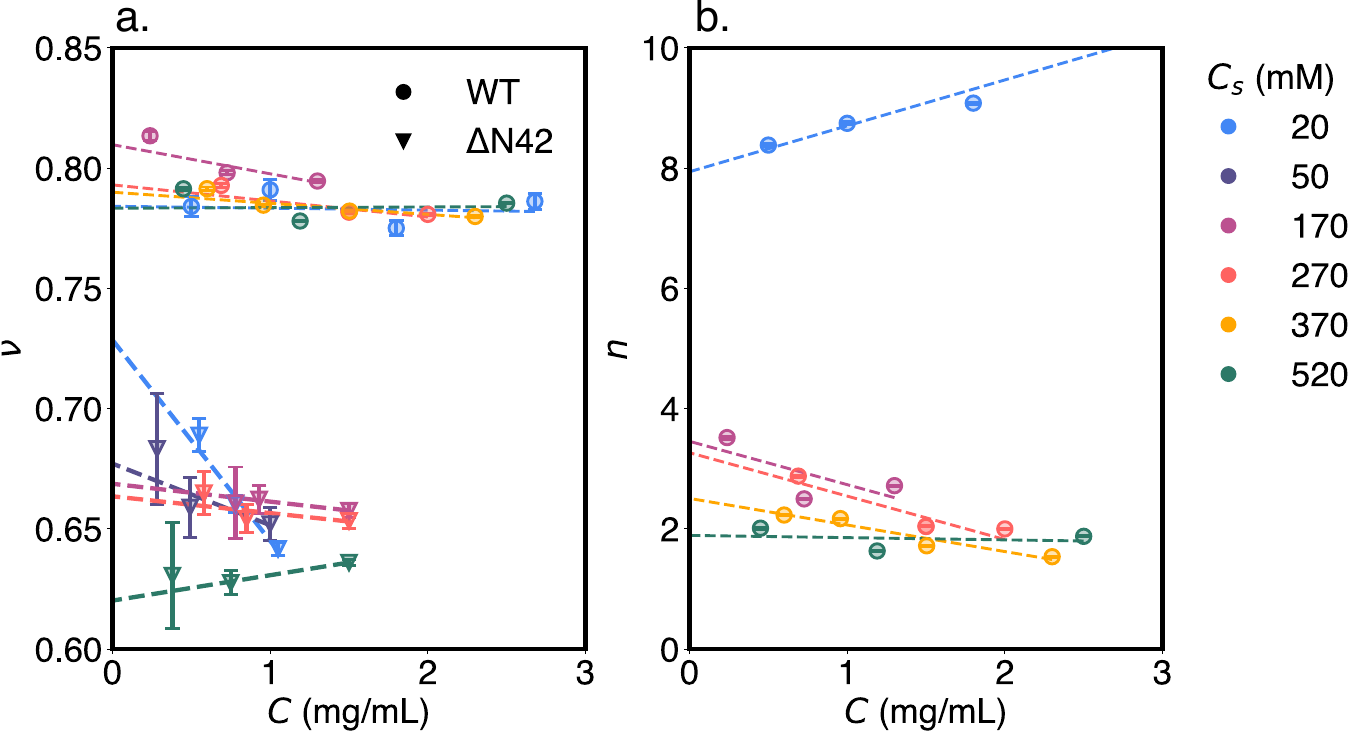}
\caption{\textbf{a.} Flory exponent ($\nu$) of WT tails
and \dn variants as a function of the concentration. \dn shows to change radically as a function of the concentration at the lowest salinities. This effect is reduced as salinity concentration $Cs$ reaches $170$mM. WT and the rest of \dn $\nu$ data shows little change as a function of the protein concentration.\textbf{b.} The core (aggregated) peptide length per polypeptide as a function of the concentrations. The large drop observed from $C_s =20$mM to $C_s =170$mM can be attributed to the shift from a dimer to a trimer. Core peptide length difference diminishes with increasing salinity, however the value still remain largely similar. }
\label{fig:sup_nu}
\end{figure*}
\begin{figure*}[hbt!]
\centering
\includegraphics[width=1\linewidth]{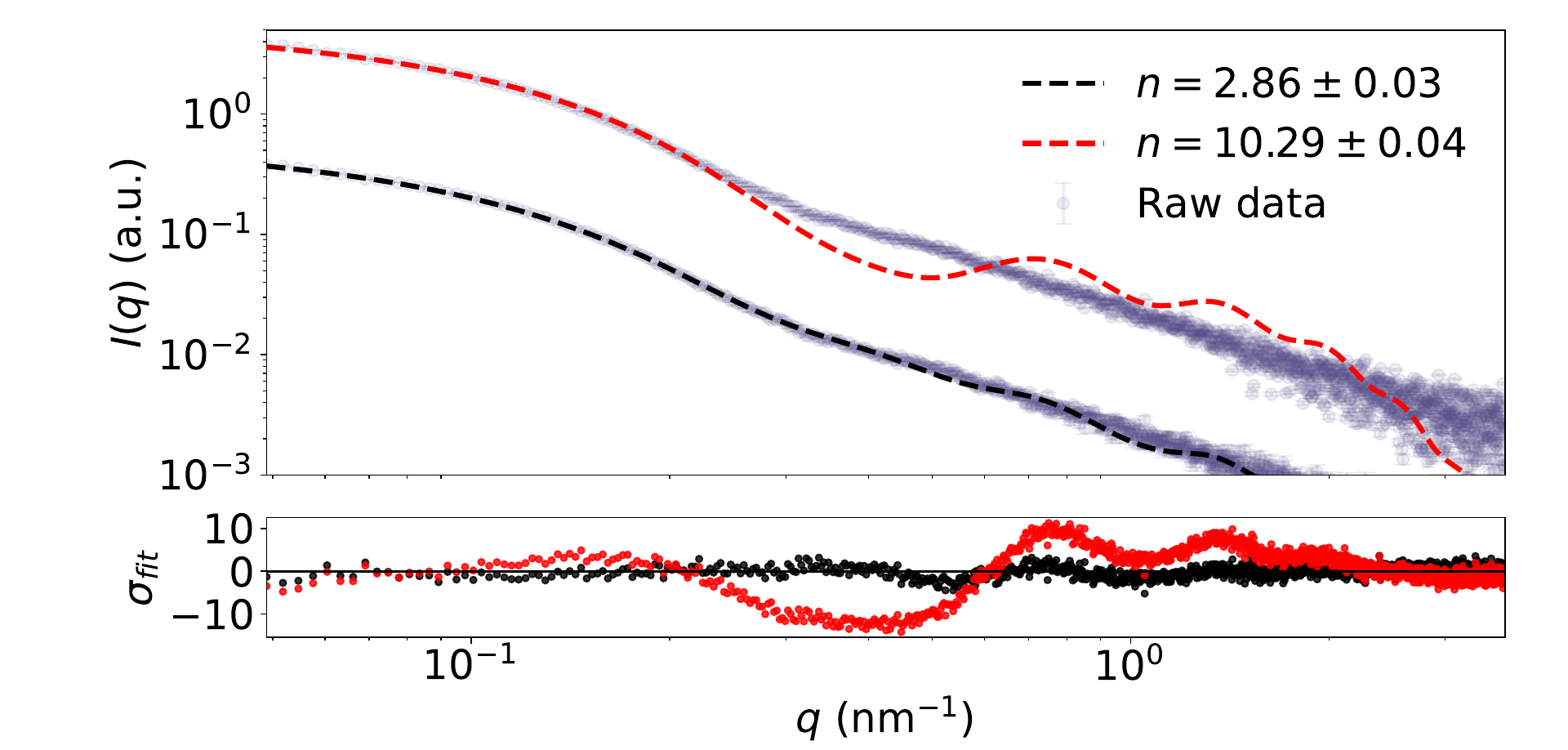}
\caption{SAXS measurement of WT and its fitting with different core residue number $n$. In fixing the core residue number to a constant value of $10$ (in red), the fitting becomes noticeably worse than when $n$ is allowed to vary (in black). Displayed data: WT in $20$ mM Tris pH=8.0, and $170$ mM NaCl at a concentration of $1.3$ mg/ml.  }
\label{fig:different_core_numbers}
\end{figure*}

\begin{figure*}[hbt!]
\centering
\includegraphics[width=1\linewidth]{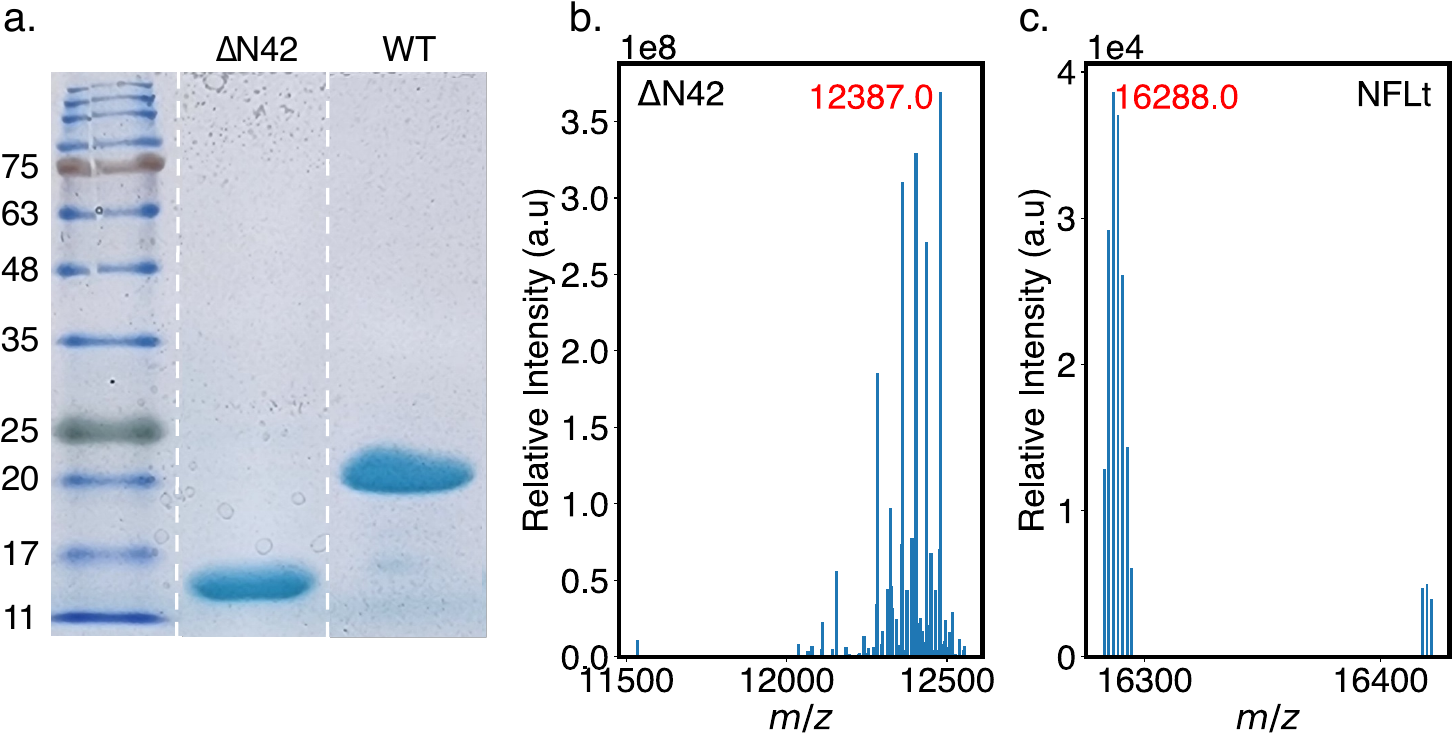}
\caption{\textbf{a.} SDS-PAGE Tris-Glycine 15\% of both \dn and NFLt (WT), showing purity above 95\%. White dashed lines indicate where image lanes were edited closer for clarity. Both show a higher molecular weight reading in the gel, which is common for IDPs.\textbf{b-c.} Deconvoluted ESI-TOF MS spectra of \dn and NFLt respectively. Theoretical molecular weight values are 12423.57 and 16233.79 for \dn and NFLt, respectively}
\label{fig:gel_sup}
\end{figure*}

\end{document}